\documentclass[prl,aps,twocolumn,showpacs,nofootinbib]{revtex4}

\usepackage{graphicx}
\usepackage{stmaryrd}
\usepackage{rotating}
\usepackage{amsmath}
\usepackage{amsfonts}
\usepackage{amssymb}
\usepackage[countmax]{subfloat}
\usepackage{dcolumn} 
\usepackage{bm} 
\usepackage{color}

\usepackage{float}
\usepackage{latexsym}

\begin{document}

\title{Majorana Edge States in Superconductor/Noncollinear Magnet Interfaces}

\author{Wei Chen$^{1,2}$ and Andreas P. Schnyder$^{2}$}

\affiliation{$^{1}$Theoretische Physik, ETH-Z\"urich, CH-8093 Z\"urich, Switzerland
\\
$^{2}$Max-Planck-Institut f$\ddot{u}$r Festk$\ddot{o}$rperforschung, Heisenbergstrasse 1, D-70569 Stuttgart, Germany}

\date{\rm\today}

\begin{abstract}

{Through $s-d$ coupling, a superconducting thin film interfaced to a noncollinear magnetic insulator inherits its magnetic order, which may induce unconventional superconductivity that hosts Majorana edge states. From the  cycloidal, helical, or (tilted) conical  magnetic order of multiferroics, or the Bloch and Neel domain walls of ferromagnetic insulators, the induced pairing is ($p_{x}+p_{y}$)-wave,
a pairing state that supports 
Majorana edge modes  without adjusting the chemical potential. In this setup, the Majorana states
can be separated over the distance of the long range magnetic order, which may reach macroscopic scale. 
A skyrmion spin texture, on the other hand, induces a ($p_{r}+ip_{\varphi}$)-wave-like state, which albeit 
nonuniform and influenced by an emergent electromagnetic field, hosts
both a bulk persistent current and a topological edge current.} 

\end{abstract}

\pacs{ 73.20.-r, 71.10.Pm, 73.21.-b, 74.45.+c}





\maketitle


{\it Introduction.-} Motivated by possible applications in non-abelian quantum computation~\cite{nayakRMP08}, the search for Majorana fermions in condensed matter systems has witnessed a boost recently~\cite{Alicea:2012em,beenakkerReview,elliott_franz_review,Stanescu_Majorana_review,schnyder_review_JPCM15}.
Indeed there has  been much effort to design and fabricate one-dimensional heterostructures 
in which  topological $p$-wave superconductivity is proximity induced~\cite{Kitaev01,Oreg10,Roman_SC_semi,Sau10,Mourik12,Das_zero_bias}.
One particularly promising proposal are chains of magnetic atoms with noncollinear spin texture 
on the surface of a conventional superconductor (SC)~\cite{Choy11,NadjPerge13,brauneckerPRL13,pientkaPRB13,klinovajaPRL13,vazifehPRL13,pientkaPRB14,rontynenPRB14, kimPRB14,brydonPRB15}.
The presence of these magnetic adatoms induces Shiba bound states~\cite{yuActaPhysSin,shibaProgTheoPhys68,rusinovJETP69},
whose low-energy physics is equivalent to a one-dimensional (1D) $p$-wave SC
with Majorana modes at its ends.
Scanning tunneling measurements of zero-bias peaks at the ends of Fe chains deposited on superconducting Pb have
been interpreted as evidence of Majorana modes~\cite{Nadj-Perge_Ferro_SC},
although no general consensus has been reached  regarding the definitive existence of Majorana  states
in these systems~\cite{brydonArxiv2014}.

Since thin films are generally easier to manufacture than adatoms, it is intriguing to ask if such 1D proposals can be generalized to two-dimensional (2D) systems, where a superconducting thin film is coupled to a noncollinear magnet. The work by Nakosai \textit{et al.} \cite{Nakosai13} first shed light on this issue.
It was found that a spiral spin texture in proximity to an $s$-wave superconductor induces
a ($p_x + p_y$)-wave state, while a skyrmion crystal spin configuration gives rise to 
($p_x + i p_y$)-wave-like pairing.
The former paring state exhibits bulk nodes and  Majorana flat-band edge states,
whereas the latter one is characterized by a full  gap with chirally dispersing Majorana edge
states that carry a quantized Hall current. In this Letter, we show that these phenomena occur in a much broader class of
superconductor/noncollinear magnet interfaces. In particular, we find that
a great variety of noncollinear magnets, including multiferroic insulators with helical, cycloidal, and (tilted) conical order, as well as magnetic domain walls of ferromagnetic (FM) insulators, interfaced with an $s$-wave superconductor induce a ($p_x + p_y$)-wave pairing state with Majorana flat bands at the boundary.
We derive the general criterion for the Majorana edge state for any wave length and direction of the noncollinear magnetic order, show that the Majorana modes occur without fine-tuning of the chemical potential,
and demonstrate that the translation invariance along the direction perpendicular to the noncollinear order greatly enhances the chance to observe Majorana states.
Furthermore, we investigate a single skyrmion spin texture coupled to an $s$-wave superconductor and 
shown that at this interface an inhomogeneous ($p_{r}+ip_{\varphi}$)-wave-like pairing is induced,
which coexists with the emerging electromagnetic field resulted from the noncoplanar spin texture.

{\it SC/multiferroic interface.-} Evidence from the $T_{c}$ reduction in magnetic oxide/SC heterostructures~\cite{Deutscher69,Hauser69} suggests that $s-d$ coupling $\Gamma{\bf S}_{i}\cdot{\boldsymbol\sigma}$ generally exists at the interface atomic layer between an SC and an insulating magnetic oxide~\cite{deGennes66}, where ${\boldsymbol\sigma}$ is the conduction electron spin and ${\bf S}_{i}$ the local moment. This leads us to consider the following model for the SC/multiferroic interface, which is the 2D generalization of the 1D proposals of Refs.~\onlinecite{Choy11,NadjPerge13},
\begin{align}  
&H=\sum_{i, \delta, \alpha}t_{i \delta}f_{i \alpha}^{\dag}f^{\ }_{i+\delta \alpha}
+t_{i \delta}^{\ast}f_{i+\delta \alpha}^{\dag}f^{\ }_{i \alpha}
-\mu\sum_{i,\alpha}f_{i \alpha}^{\dag}f_{i \alpha}
\nonumber \\
&+\sum_{i,\alpha,\beta}\left({\bf B}_{i}\cdot{\boldsymbol\sigma}\right)_{\alpha \beta}f_{i \alpha}^{\dag}f^{\ }_{i \beta}
+\sum_{i}\Delta_{0}\left(f_{i \uparrow}^{\dag}f_{i \downarrow}^{\dag}
+f^{\ }_{i \downarrow}f^{\ }_{i \uparrow}\right),\;\;\;\;
\label{1D_Hamiltonian}
\end{align} 
where $i=\left\{i_{x},i_{y}\right\}$ is the site index, ${\hat{\boldsymbol\delta}}=\left\{{\hat{\bf a}},{\hat{\bf b}}\right\}$ is the planar unit vector, and $\alpha$ is the spin index.  In the absence of spin-orbit interaction, the majority of the noncollinear order discovered in multiferroics, for instance in perovskite rare earth manganites~\cite{Yamasaki07,Yamasaki08,Kimura08,Murakawa08}, can be generically described by the conical order
\begin{eqnarray}
{\bf B}_{i}=(B_{i}^{x},B_{i}^{y},B_{i}^{z})=\left(B_{\parallel}\sin\theta_{i},B_{\perp},B_{\parallel}\cos\theta_{i}\right)\; ,
\label{Bfield_general}
\end{eqnarray}
as far as their effect on the SC is concerned, since the choice of coordinate for ${\boldsymbol\sigma}$ is arbitrary. For instance, cycloidal and helical order are equivalent by trivially exchange two components of ${\boldsymbol\sigma}$, and conical orders with any tilting angle are equivalent. Here $B_{\parallel}=\Gamma|{\bf S}_{i\parallel}|$ and $B_{\perp}=\Gamma|{\bf S}_{i\perp}|$ are the planar and out-of-plane components of the local moment, and the planar angle $\theta_{i}={\bf Q}\cdot{\bf r}_{i}$ is determined by the spiral wave vector ${\bf Q}$ and the planar position ${\bf r}_{i}$. The cycloidal order is the case when $B_{\perp}{\hat{\bf y}}=0$. Note that a three-dimensional conical order projected to a surface cleaved at any direction is still that described by Eq.~(\ref{Bfield_general}), so our formalism is applicable to a thin film or the surface of a single crystal multiferroic in any crystalline orientation, assuming no lattice mismatch with the SC and a constant $|{\bf S}_{i}|=|{\bf S}|$.

We perform two consecutive rotations to align the ${\bf B}_{i}$ field along $\sigma^{z}$,
\begin{eqnarray}
\left(
\begin{array}{l}
f_{i\uparrow} \\
f_{i\downarrow}
\end{array}
\right)
&=&\left(
\begin{array}{ll}
\cos\frac{\theta_{i}}{2} & -\sin\frac{\theta_{i}}{2} \\
\sin\frac{\theta_{i}}{2} & \cos\frac{\theta_{i}}{2}
\end{array}
\right)
\left(
\begin{array}{ll}
\cos\frac{\gamma}{2} & i\sin\frac{\gamma}{2} \\
i\sin\frac{\gamma}{2} & \cos\frac{\gamma}{2}
\end{array}
\right)
\left(
\begin{array}{l}
g_{i\uparrow} \\
g_{i\downarrow}
\end{array}
\right)
\nonumber \\
&=&U_{i}
\left(
\begin{array}{l}
g_{i\uparrow} \\
g_{i\downarrow}
\end{array}
\right) ,
\label{SU2_rotation}
\end{eqnarray} 
where $\sin\gamma=B_{\perp}/B_{0}$ and $B_{0}=\sqrt{B_{\parallel}^{2}+B_{\perp}^{2}}=|\Gamma{\bf S}|$, under which one obtains
\begin{eqnarray}
&&
H=\sum_{i,\delta,\alpha,\beta}t_{i\delta}\Omega_{i\delta\alpha\beta}g_{i\alpha}^{\dag}g^{\ }_{i+\delta\beta}
+t_{i\delta}^{\ast}\left(\Omega_{i\delta}^{\ast}\right)_{\beta\alpha}g_{i+\delta\alpha}^{\dag}g^{\ }_{i\beta}
\nonumber \\
&& 
+\sum_{i, \alpha, \beta}\left(B\sigma_{\alpha\beta}^{z}-\mu I_{\alpha\beta}\right)g_{i\alpha}^{\dag}g^{\ }_{i\beta}
+\sum_{i}\Delta_{0}\left(g_{i\uparrow}^{\dag}g_{i\downarrow}^{\dag}
+g^{\ }_{i\downarrow}g^{\ }_{i\uparrow}\right),
\nonumber \\
&&\Omega_{i\delta}=U_{i}^{\dag}U^{\ }_{i+\delta}=
\left(
\begin{array}{ll}
\alpha_{i\delta} & -\beta_{i\delta}^{\ast} \\
\beta_{i\delta} & \alpha_{i\delta}^{\ast}
\end{array}
\right)\;,
\nonumber \\
&&\alpha_{i\delta}\equiv\alpha_{\delta}=\cos\frac{{\bf Q}\cdot {\boldsymbol\delta}}{2}-i\sin\gamma\sin\frac{{\bf Q}\cdot{\boldsymbol\delta}}{2}\;,
\nonumber \\
&&\beta_{i\delta}\equiv\beta_{\delta}=\cos\gamma\sin\frac{{\bf Q}\cdot {\boldsymbol\delta}}{2}\;.
\label{Hamiltonian_gi_basis}
\end{eqnarray}
In what follows, we consider hopping to be real and isotropic $t_{i\delta}=t_{i\delta}^{\ast}=t$.

In the limit $B_{0}\approx|\mu|\gg \left\{t,\Delta_{0}\right\}$ with $\mu<0$, one can construct an effective low energy theory for the spin species near the Fermi level, which is the spin down band. This is done by introducing a unitary transformation $H^{\prime}=e^{-iS}He^{iS}$ to eliminate the spin mixing terms order by order~\cite{Choy11}. At first order, the pairing part 
\begin{eqnarray}
H_{\textrm{eff},\Delta}=\sum_{i, \delta}\left[\left(\frac{1}{B}-\frac{1}{\mu}\right)\Delta_{0}t\beta_{i\delta}^{\ast}g_{i\downarrow}g_{i+\delta\downarrow}+\textrm{h.c.}\right]
\label{low_energy_Heff}
\end{eqnarray}
resembles a spinless $p$-wave superconductor with anisotropic nearest-neighbor and next-nearest-neighbor hopping. From Eq.~(\ref{Hamiltonian_gi_basis}) one has $\beta_{i\delta}^{\ast}=\cos\gamma\sin{\bf Q}\cdot{\boldsymbol\delta}/2$, so the induced pairing is of ($p_{x}+p_{y}$)-wave symmetry, with the magnitude of the gap determined by the wave length of the planar component of the conical order.

To derive the criterion for the appearance of Majorana edge states,
we introduce Majorana fermions 
$g_{i\sigma}=\frac{1}{2}\left(b_{i1\sigma}+ib_{i2\sigma}\right)$, $g_{i\sigma}^{\dag}=\frac{1}{2}\left(b_{i1\sigma}-ib_{i2\sigma}\right)$ with $\left\{b_{im\alpha},b_{i^{\prime}m^{\prime}\beta}\right\}=2\delta_{ii^{\prime}}\delta_{mm^{\prime}}\delta_{\alpha\beta}$, 
and express the Hamiltonian $H=\left(i/4\right)\sum_{\bf q}b_{\bf q}^{\dag}A(\bf q)b_{\bf -q}$ in terms of the basis $b_{\bf q}^{\dag}=\left(b_{{\bf q}1\uparrow},b_{{\bf q}1\downarrow},b_{{\bf q}2\uparrow},b_{{\bf q}2\downarrow}\right)$\cite{NadjPerge13}. We choose open boundary condition (OBC) along ${\hat{\bf x}}$ and periodic one (PBC) along ${\hat{\bf y}}$, such that $-\pi<q_{y}<\pi$ is a good quantum number. At a particular $q_{y}$, only when $q_{x}$ satisfies 
\begin{eqnarray}
\beta_{a}\sin q_{x}+\beta_{b}\sin q_{y}=0\;,
\end{eqnarray}
is the Majorana Hamiltonian $A(q_{x},q_{y})$ skew symmetric. The two solutions $q_{x1}$ and $q_{x2}$ are the high symmetry points at which the Pfaffian is calculated. The topological index~\cite{NadjPerge13} is 
now a function of $q_{y}$
\begin{eqnarray}
M(q_{y})={\rm Sign}\left(\mathrm{Pf}\left[A(q_{x1},q_{y})\right]\right)
{\rm Sign}\left(\mathrm{Pf}\left[A(q_{x2},q_{y})\right]\right)\;.\;\;\;
\label{judging_topo_index}
\end{eqnarray}
When $M(q_{y})=-1$, or equivalently
\begin{eqnarray}
&&\sqrt{\Delta_{0}^{2}+\left(|\mu-2t\overline{\alpha}_{b}\cos q_{y}|+|2t\overline{\alpha}_{a}\cos q_{x1}|\right)^{2}}>B
\nonumber \\
&&>\sqrt{\Delta_{0}^{2}+\left(|\mu-2t\overline{\alpha}_{b}\cos q_{y}|-|2t\overline{\alpha}_{a}\cos q_{x1}|\right)^{2}}\;,
\label{topo_index}
\end{eqnarray}
where $\overline{\alpha}_{\delta}=\left(\alpha_{\delta}+\alpha_{\delta}^{\ast}\right)/2$, the Majorana edge state with momentum $q_{y}$ appears. Setting $\overline{\alpha}_{b}=0$ and $q_{x1}=0$ recovers the well known 1D result~\cite{NadjPerge13}.

\begin{figure}[ht]
\begin{center}
\includegraphics[clip=true,width=0.98\columnwidth]{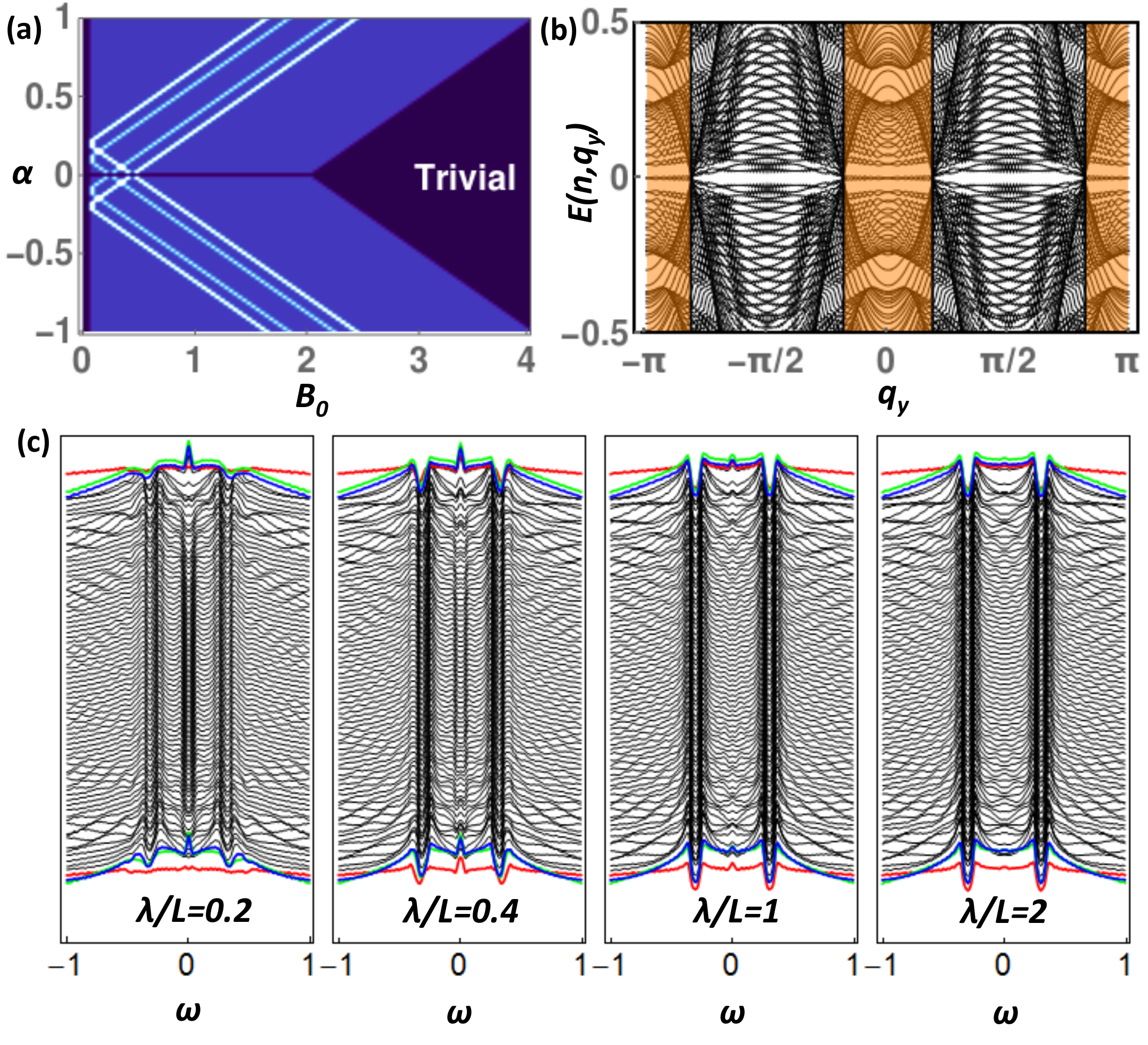}
\caption{ (color online) (a) The topologically nontrivial phase (blue region) of the 2D model Eq.~(\ref{1D_Hamiltonian}) with $\mu=0$ as a function of $s-d$ coupling $B_{0}$ and $\alpha\equiv {\overline\alpha}_{a}$ for $(1,0)$ spiral. 
For comparison, the nontrivial phase of 1D model with $\mu=-0.2$ and $-0.4$ are shown as the region inside light blue dots and white dots, respectively. (b) Energy levels of a system with size $L=80a$ subject to a spiral of wave length $\lambda=2\pi/|{\bf Q}|=16a$ at $B_{0}=0.3$. The orange regions are the $q_{y}$'s at which the topological criterion Eq.~(\ref{topo_index}) is satisfied, where one also sees the Majorana zero energy edge state. (c) the LDOS along ${\hat{\bf x}}$ for different values of $\lambda/L$. Red, green, and blue lines are the first, second, and third site away from the edge. The $\lambda/L=2$ case represents a Neel or Bloch domain wall in a ferromagnetic insulator. Other parameters are $t=-1$, $\Delta=0.05$.  } 
\label{fig:SSC_spiral_example}
\end{center}
\end{figure}

Equation (\ref{topo_index}) is the general criterion for the Majorana state to appear at momentum $q_y$ for any given spiral or conical order.  It is supported by numerically solving the Bogoliubov-de Gennes (BdG) equation with the boundary conditions we choose. A spin-generalized Bogoliubov transformation $g_{i\sigma}=\sum_{n,\alpha}u_{in\sigma\alpha}\gamma_{n\alpha}
+v_{in\sigma\alpha}^{\ast}\gamma_{n\alpha}^{\dag}$ is introduced to diagonalize Eq.~(\ref{Hamiltonian_gi_basis}). Figure \ref{fig:SSC_spiral_example}(b) shows a typical dispersion $E(n,q_{y})$, which display Majorana zero-energy states in the $q_{y}$'s that satisfy Eq.~(\ref{topo_index}). Note that Equation (\ref{topo_index}) can be satisfied even if $\mu=0$, so adjusting chemical potential is generally not needed. Consequently, the edge states can occur in an isolated sample without attaching any leads. The $2t\overline{\alpha}_{b}\cos q_{y}$ factor greatly enlarges the number of $q_{y}$'s that can satisfy Eq.~(\ref{topo_index}), hence increases the chance to observe Majorana Fermions, as one can see from the phase diagram shown in Fig.~\ref{fig:SSC_spiral_example}(a) that has much larger topologically nontrivial phase in the $B_{0}$-${\overline\alpha}$ space than the 1D cases, where the (weak) topological phase~\cite{Sedlmayr15} is judged by whether any $q_y$ satisfies Eq.~(\ref{topo_index})  at a given $(B_{0},{\overline\alpha})$.

The localized Majorana edge states can be seen as the zero bias peaks (ZBPs) in the local density of states (LDOS) along ${\hat{\bf x}}$ direction, as plotted in Fig.~\ref{fig:SSC_spiral_example}(c). Three gap-like features show up in the LDOS. The two symmetric in $\omega$ come from the bulk gap $\Delta$ whose position is shifted by the $s-d$ coupling $\pm B_{0}$ as its effect is similar to a magnetic field, and typically has a magnitude~\cite{Kajiwara10} $0.01\sim 1$eV so $B_{0}>\Delta$. The gap-like feature near zero energy represents the induced ($p_{x}+p_{y}$)-wave gap in Eq.~(\ref{low_energy_Heff}). The weight of ZBP decreases as the spiral wave length $\lambda=2\pi/|{\bf Q}|$ increases. The $\lambda/L=2$ case represents a Neel or Bloch magnetic domain wall joining two regions of opposite spin orientations in a ferromagnetic insulator, since it can be viewed as a spiral with half a wave length, and the interface to a SC can be described by Eqs.~(\ref{1D_Hamiltonian}) to (\ref{topo_index}). In such case, the ZBP is small but still discernable.

For a thin film with finite thickness but the $s-d$ coupling only at the interface atomic layer, the Majorana state extends over few layers away from the interface, so one may need a SC film of few atomic layers thickness to observe the Majoranas by any surface probe such as scanning tunneling microscope (STM). Even if the multiferroic contains domains of different spiral chirality, or the spin texture is not perfectly periodic, the Majorana edge state still exits at the edge and the boundary between domains. The large single domain of multiferroics, currently of mm size~\cite{Johnson13}, may help to separate the Majorana fermions over a distance of macroscopic scale.

\begin{figure}[t!]
\begin{center}
\includegraphics[clip=true,width=0.98\columnwidth]{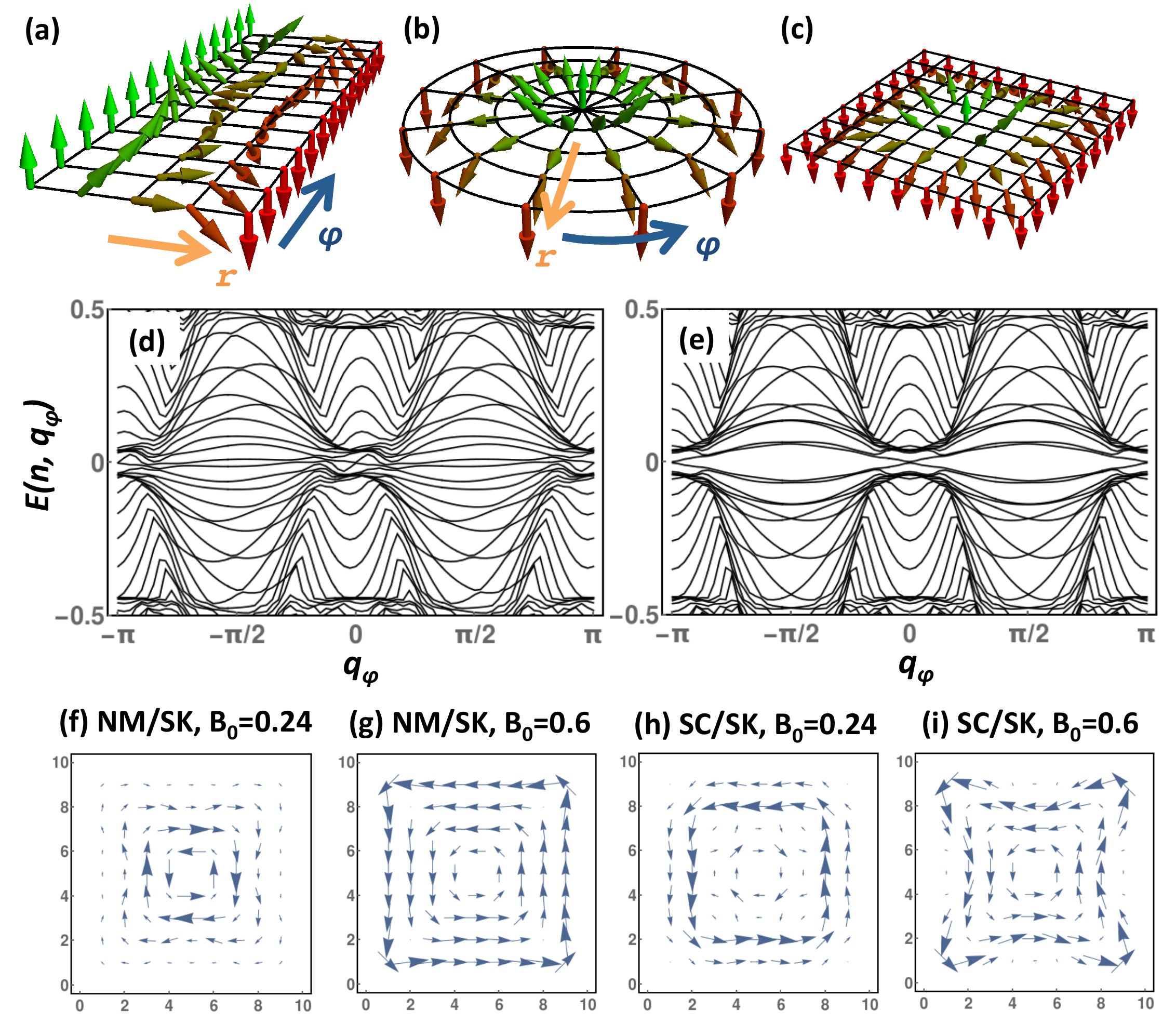}
\caption{ (color online) (a) The spin texture of the pedagogical model, which is closely related to a single nonchiral skyrmion in (b) a polar lattice and (c)  a square lattice. (d) The dispersion of the pedagogical model with OBC in ${\hat{\bf r}}$ and PBC in ${\hat{\boldsymbol \varphi}}$, with $N_{r}=N_{\varphi}=40$, and (e) if the emerging EM field is manually turned off, at $B_{0}=0.24$. The spontaneous current pattern at the interface to a single nonchiral skyrmion of size $9\times 9$ is shown in (f), (g) for the normal state ($\Delta=0$) and (h), (i) for the SC state. Parameters are $t=-1$, $\Delta=0.2$, $\mu=-0.1$.  } 
\label{fig:SCskyrmion_result}
\end{center}
\end{figure}

{\it SC/skyrmion interface.-} Skyrmion spin textures have been observed in thin film insulating multiferroics~\cite{Seki12,Seki12_2} at temperatures approaching the typical SC transition temperature, with a small magnetic field that presumably has negligible effect on the SC. To gain more understanding about the SC/skyrmion insulator interface, disregarding external magnetic fields, we first consider a closely related pedagogical model defined on a square lattice, whose low-energy sector can be studied analytically. The spin texture of this model is that shown in Fig.~\ref{fig:SCskyrmion_result}(a), yielding a magnetic field on its interface to an SC
\begin{eqnarray}
{\bf B}_{i}=B\left(\sin\theta_{i}\cos\varphi_{i},\sin\theta_{i}\sin\varphi_{i},\cos\theta_{i}\right)
\label{skyrmion_B}
\end{eqnarray}
at position $(r_{i},\varphi_{i})$ on the square lattice, where $\theta_{i}=\pi r_{i}/R$ and $R$ is the width in ${\hat{\bf r}}$ direction.
We assume OBC along the ${\hat{\bf r}}$ direction and PBC along the ${\hat{\boldsymbol\varphi}}$ direction.
The Hamiltonian is described by Eq.~(\ref{1D_Hamiltonian}) with ${\hat{\boldsymbol\delta}}=\left\{{\hat{\bf r}},{\hat{\boldsymbol\varphi}}\right\}$. To align the spin texture along $\sigma^{z}$, one performs the rotation in Eq.~(\ref{SU2_rotation}) with $U_{i}$ defined by
\begin{eqnarray}
U_{i}
=\left(
\begin{array}{ll}
\cos\frac{\theta_{i}}{2} & -\sin\frac{\theta_{i}}{2}e^{-i\varphi_{i}} \\
\sin\frac{\theta_{i}}{2}e^{i\varphi_{i}} & \cos\frac{\theta_{i}}{2}
\end{array}
\right)\;,
\label{skyrmion_rotation}
\end{eqnarray}
which yields Eq.~(\ref{Hamiltonian_gi_basis}) with 
\begin{eqnarray}
\alpha_{i\varphi}&=&1+2ie^{i\pi/N_{\varphi}}\sin^{2}\left(\frac{\theta_{i}}{2}\right)\sin\frac{\pi}{N_{\varphi}}\;,
\;\;\alpha_{ir}=\cos\frac{\pi}{N_{r}}\;,
\nonumber \\
\beta_{i\varphi}&=&i\sin\theta_{i}e^{i\left(\varphi_{i}+\varphi_{i+\varphi}\right)/2}\sin\frac{\pi}{N_{\varphi}}
\equiv e^{i\left(\varphi_{i}+\varphi_{i+\varphi}\right)/2}\tilde{\beta}_{i\varphi}\;,
\nonumber \\
\beta_{ir}&=&e^{i\varphi_{i}}\sin\frac{\pi}{N_{r}}\equiv e^{i\varphi_{i}}\tilde{\beta}_{ir}\;,
\label{skyrmion_alpha_beta}
\end{eqnarray}
where $N_{r}$ and $N_{\varphi}$ are the number of sites in each direction. After gauging away the extra phase $\beta_{i\delta}\rightarrow\tilde{\beta}_{i\delta}$ by $e^{-i\varphi_{i}/2}g_{i\downarrow}\rightarrow g_{i\downarrow}$ and $e^{i\varphi_{i}/2}g_{i\uparrow}\rightarrow g_{i\uparrow}$, the Hamiltonian is translationally invariant along ${\hat{\boldsymbol\varphi}}$ but not along ${\hat{\bf r}}$ because $\tilde{\beta}_{i\varphi}\propto \sin\theta_{i}=\sin\left(\pi r_{i}/R\right)$. In the $B\approx|\mu|\gg\left\{t_{\delta},\Delta_{0}\right\}$ limit, using Eq.~(\ref{low_energy_Heff}) of the low-energy effective theory, the induced gap along ${\hat{\bf r}}$ and ${\hat{\boldsymbol\varphi}}$ are proportional to $\tilde{\beta}_{ir}^{\ast}$ and $\tilde{\beta}_{i\varphi}^{\ast}$, and therefore of $(p_{r}+ip_{\varphi})$-wave-like symmetry.


The spin-conserved $\alpha_{i\delta}t_{i\delta}$ and spin-flip $\beta_{i\delta}t_{i\delta}$ hopping in the $g_{i\sigma}$ basis contain an emergent electromagnetic (EM) field~\cite{Schulz12,Nagaosa13} coming from the spatial dependence of the unitary transformation~\cite{Zhang09}. This becomes evident in the continuous limit $\theta_{i}\rightarrow\theta$, $\varphi_{i}\rightarrow\varphi$, $U_{i}\rightarrow U$, and using $\int_{0}^{a_{r}}dr\partial_{r}\theta=2\pi/N_{r}\ll 1$ and $\int_{0}^{a_{\varphi}}d\varphi\partial_{\varphi}\varphi=2\pi/N_{\varphi}\ll 1$. The $\left\{\alpha_{\delta},\beta_{\delta}\right\}$ contain the phase gained over one lattice constant
\begin{eqnarray}
\Omega_{\delta}=
\left(
\begin{array}{ll}
\alpha_{\delta} & -\beta_{\delta}^{\ast} \\
\beta_{\delta} & \alpha_{\delta}^{\ast}
\end{array}
\right)=e^{-iq\int_{0}^{a_{\delta}}d\delta A_{\delta}}\;,
\label{emerging_gauge}
\end{eqnarray}
where $A_{\delta}=i\left(\hbar/q\right)U^{\dag}\partial_{\delta}U$. The factor of $i$ difference between $A_{r}$ and $A_{\varphi}$ eventually leads to the induced $(p_{r}+ip_{\varphi})$-wave-like gap. Figure \ref{fig:SCskyrmion_result}(d) shows the dispersion of this pedagogical model, and Fig.~\ref{fig:SCskyrmion_result}(e) shows the dispersion when the emergent EM field is manually turned off by setting $\alpha_{i\varphi}=1$ in Eq.~(\ref{skyrmion_alpha_beta}). Without the emergent EM field, the dispersive edge bands expected for the induced ($p_{r}+ip_{\varphi}$)-wave-like gap are evident, whereas in the presence of it the bulk gap is diminished, although the trace of edge bands can still be seen in som cases (compare $q_{\varphi}\approx 0$ and $E(n,q_{\varphi})\approx 0$ regions in Fig.~\ref{fig:SCskyrmion_result}(d) and (e)).


The relevance of this pedagogical model is made clear by shrinking the spins at the $r_{i}=0$ edge (green arrows in Fig.~\ref{fig:SCskyrmion_result}(a)) into one single spin, which results in a nonchiral skyrmion on a polar lattice shown in Fig.~\ref{fig:SCskyrmion_result}(b), with the same interface magnetic field described by Eq.~(\ref{skyrmion_B}). Certainly these two lattices cannot be mapped to each other exactly, but their low energy sectors display similar features.

The pedagogical model indicates that the SC/skyrmion interface hosts a complex interplay between (i) the $s$-wave gap, (ii) the induced ($p_{r}+ip_{\varphi}$)-wave-like gap, (iii) the emergent EM field, and (iv) the $s-d$ coupling $B_{0}$. Motivated by the STM generated single skyrmion~\cite{Romming13} (although with an external magnetic field), we proceed to study the SC/skyrmion interface on a single open square (SC/SK), whose spin texture ${\bf B}_{i}$ at position ${\bf r}_{i}=\left(x_{i},y_{i}\right)=|{\bf r}_{i}|\left(\cos\varphi_{i},\sin\varphi_{i}\right)$, as shown in Fig.~\ref{fig:SCskyrmion_result}(c), is that described by Eq.~(\ref{skyrmion_B}), but with $\theta_{i}=\pi|{\bf r}_{i}|/R({\bf r}_{i})$. Here, $R({\bf r}_{i})$ is the length of the straight line that passes through ${\bf r}_{i}$ connecting the center of the square with the edge. The spontaneous current at site $i$ can be calculated from Eq.~(\ref{1D_Hamiltonian}) by
\begin{eqnarray}
{\bf J}_{i}=-i\sum_{\sigma\delta}t_{\delta}{\boldsymbol\delta}\langle f_{i+\delta\sigma}^{\dag}f_{i\sigma}\rangle\;.
\end{eqnarray}
As shown in Fig.~\ref{fig:SCskyrmion_result}(f) and (g), in the normal state (NM/SK) there is a persistent current whose  vorticity strongly depends on the emergent EM field, the $s-d$ coupling $B_{0}$, and finite chemical potential. Because the persistent current also has edge component, the topological edge current alone is hard to be quantified in the SC state from the pattern shown in Fig.~\ref{fig:SCskyrmion_result}(h) and (i). At $B_{0}\gg \Delta$, the current pattern in the SC state recovers that of the NM state, as can be seen by comparing Fig.~\ref{fig:SCskyrmion_result}(g) and (i). For the chiral skyrmion seen in most experiments, the form of $B^{x}$ and $B^{y}$ in Eq.~(\ref{skyrmion_B}) are exchanged, which gives the same result as $\sigma^{x}$ and $\sigma^{y}$ can be trivially exchanged for spin-independent quantities such as ${\bf J}_{i}$. These results suggest a vortex-like state at the NM/skyrmion lattice or SC/skyrmion lattice interface, whose vorticity depends on material properties.




In summary, we propose that a broad class of noncollinear magnetic orders, including a large part of those discovered in multiferroic insulators, and the Bloch and Neel domain walls in collinear magnetic insulators, can be used to practically generate Majorana edge states at their interface to a conventional SC. The advantages of these systems include a much larger parameter space to stabilize the edge states compared to 1D proposals, the longer range magnetic order may help to separate the edge states over a macroscopic distance, and adjusting chemical potential is not necessary. The proximity to a skyrmion induces an inhomogeneous ($p_{r}+ip_{\varphi}$)-wave-like pairing in the SC under the influence of an emergent electromagnetic field, and consequently a vortex-like state that features both a bulk persistent current and an edge current.


We thank P. W. Brouwer, Y.-H. Liu, and F. von Oppen for stimulating discussions.

\bibliography{refs_v1}

\begin{thebibliography}{44}
\expandafter\ifx\csname natexlab\endcsname\relax\def\natexlab#1{#1}\fi
\expandafter\ifx\csname bibnamefont\endcsname\relax
  \def\bibnamefont#1{#1}\fi
\expandafter\ifx\csname bibfnamefont\endcsname\relax
  \def\bibfnamefont#1{#1}\fi
\expandafter\ifx\csname citenamefont\endcsname\relax
  \def\citenamefont#1{#1}\fi
\expandafter\ifx\csname url\endcsname\relax
  \def\url#1{\texttt{#1}}\fi
\expandafter\ifx\csname urlprefix\endcsname\relax\def\urlprefix{URL }\fi
\providecommand{\bibinfo}[2]{#2}
\providecommand{\eprint}[2][]{\url{#2}}

\bibitem[{\citenamefont{Nayak et~al.}(2008)\citenamefont{Nayak, Simon, Stern,
  Freedman, and Das~Sarma}}]{nayakRMP08}
\bibinfo{author}{\bibfnamefont{C.}~\bibnamefont{Nayak}},
  \bibinfo{author}{\bibfnamefont{S.~H.} \bibnamefont{Simon}},
  \bibinfo{author}{\bibfnamefont{A.}~\bibnamefont{Stern}},
  \bibinfo{author}{\bibfnamefont{M.}~\bibnamefont{Freedman}}, \bibnamefont{and}
  \bibinfo{author}{\bibfnamefont{S.}~\bibnamefont{Das~Sarma}},
  \bibinfo{journal}{Rev. Mod. Phys.} \textbf{\bibinfo{volume}{80}},
  \bibinfo{pages}{1083} (\bibinfo{year}{2008}).

\bibitem[{\citenamefont{Alicea}(2012)}]{Alicea:2012em}
\bibinfo{author}{\bibfnamefont{J.}~\bibnamefont{Alicea}},
  \bibinfo{journal}{Reports on Progress in Physics}
  \textbf{\bibinfo{volume}{75}}, \bibinfo{pages}{076501}
  (\bibinfo{year}{2012}).

\bibitem[{\citenamefont{Beenakker}(2013)}]{beenakkerReview}
\bibinfo{author}{\bibfnamefont{C.}~\bibnamefont{Beenakker}},
  \bibinfo{journal}{Annual Review of Condensed Matter Physics}
  \textbf{\bibinfo{volume}{4}}, \bibinfo{pages}{113} (\bibinfo{year}{2013}).

\bibitem[{\citenamefont{Elliott and Franz}(2015)}]{elliott_franz_review}
\bibinfo{author}{\bibfnamefont{S.~R.} \bibnamefont{Elliott}} \bibnamefont{and}
  \bibinfo{author}{\bibfnamefont{M.}~\bibnamefont{Franz}},
  \bibinfo{journal}{Rev. Mod. Phys.} \textbf{\bibinfo{volume}{87}},
  \bibinfo{pages}{137} (\bibinfo{year}{2015}).

\bibitem[{\citenamefont{Stanescu and Tewari}(2013)}]{Stanescu_Majorana_review}
\bibinfo{author}{\bibfnamefont{T.~D.} \bibnamefont{Stanescu}} \bibnamefont{and}
  \bibinfo{author}{\bibfnamefont{S.}~\bibnamefont{Tewari}},
  \bibinfo{journal}{J. Phys. Condens. Matter} \textbf{\bibinfo{volume}{25}},
  \bibinfo{pages}{233201} (\bibinfo{year}{2013}).

\bibitem[{\citenamefont{{Schnyder} and
  {Brydon}}(2015)}]{schnyder_review_JPCM15}
\bibinfo{author}{\bibfnamefont{A.~P.} \bibnamefont{{Schnyder}}}
  \bibnamefont{and} \bibinfo{author}{\bibfnamefont{P.~M.~R.}
  \bibnamefont{{Brydon}}}, \bibinfo{journal}{ArXiv e-prints}
  (\bibinfo{year}{2015}), \eprint{1502.03746}.

\bibitem[{\citenamefont{{Kitaev}}(2001)}]{Kitaev01}
\bibinfo{author}{\bibfnamefont{A.}~\bibnamefont{{Kitaev}}},
  \bibinfo{journal}{Physics Uspekhi} \textbf{\bibinfo{volume}{44}},
  \bibinfo{pages}{131} (\bibinfo{year}{2001}).

\bibitem[{\citenamefont{Oreg et~al.}(2010)\citenamefont{Oreg, Refael, and von
  Oppen}}]{Oreg10}
\bibinfo{author}{\bibfnamefont{Y.}~\bibnamefont{Oreg}},
  \bibinfo{author}{\bibfnamefont{G.}~\bibnamefont{Refael}}, \bibnamefont{and}
  \bibinfo{author}{\bibfnamefont{F.}~\bibnamefont{von Oppen}},
  \bibinfo{journal}{Phys. Rev. Lett.} \textbf{\bibinfo{volume}{105}},
  \bibinfo{pages}{177002} (\bibinfo{year}{2010}).

\bibitem[{\citenamefont{Lutchyn et~al.}(2010)\citenamefont{Lutchyn, Sau, and
  Das~Sarma}}]{Roman_SC_semi}
\bibinfo{author}{\bibfnamefont{R.~M.} \bibnamefont{Lutchyn}},
  \bibinfo{author}{\bibfnamefont{J.~D.} \bibnamefont{Sau}}, \bibnamefont{and}
  \bibinfo{author}{\bibfnamefont{S.}~\bibnamefont{Das~Sarma}},
  \bibinfo{journal}{Phys. Rev. Lett.} \textbf{\bibinfo{volume}{105}},
  \bibinfo{pages}{077001} (\bibinfo{year}{2010}).

\bibitem[{\citenamefont{Sau et~al.}(2010)\citenamefont{Sau, Lutchyn, Tewari,
  and Das~Sarma}}]{Sau10}
\bibinfo{author}{\bibfnamefont{J.~D.} \bibnamefont{Sau}},
  \bibinfo{author}{\bibfnamefont{R.~M.} \bibnamefont{Lutchyn}},
  \bibinfo{author}{\bibfnamefont{S.}~\bibnamefont{Tewari}}, \bibnamefont{and}
  \bibinfo{author}{\bibfnamefont{S.}~\bibnamefont{Das~Sarma}},
  \bibinfo{journal}{Phys. Rev. Lett.} \textbf{\bibinfo{volume}{104}},
  \bibinfo{pages}{040502} (\bibinfo{year}{2010}).

\bibitem[{\citenamefont{Mourik et~al.}(2012)\citenamefont{Mourik, Zuo, Frolov,
  Plissard, Bakkers, and Kouwenhoven}}]{Mourik12}
\bibinfo{author}{\bibfnamefont{V.}~\bibnamefont{Mourik}},
  \bibinfo{author}{\bibfnamefont{K.}~\bibnamefont{Zuo}},
  \bibinfo{author}{\bibfnamefont{S.~M.} \bibnamefont{Frolov}},
  \bibinfo{author}{\bibfnamefont{S.~R.} \bibnamefont{Plissard}},
  \bibinfo{author}{\bibfnamefont{E.~P. A.~M.} \bibnamefont{Bakkers}},
  \bibnamefont{and} \bibinfo{author}{\bibfnamefont{L.~P.}
  \bibnamefont{Kouwenhoven}}, \bibinfo{journal}{Science}
  \textbf{\bibinfo{volume}{336}}, \bibinfo{pages}{1003} (\bibinfo{year}{2012}).

\bibitem[{\citenamefont{Das et~al.}(2012)\citenamefont{Das, Ronen, Most, Oreg,
  Heiblum, and Shtrikman}}]{Das_zero_bias}
\bibinfo{author}{\bibfnamefont{A.}~\bibnamefont{Das}},
  \bibinfo{author}{\bibfnamefont{Y.}~\bibnamefont{Ronen}},
  \bibinfo{author}{\bibfnamefont{Y.}~\bibnamefont{Most}},
  \bibinfo{author}{\bibfnamefont{Y.}~\bibnamefont{Oreg}},
  \bibinfo{author}{\bibfnamefont{M.}~\bibnamefont{Heiblum}}, \bibnamefont{and}
  \bibinfo{author}{\bibfnamefont{H.}~\bibnamefont{Shtrikman}},
  \bibinfo{journal}{Nat. Phys.} \textbf{\bibinfo{volume}{8}},
  \bibinfo{pages}{887} (\bibinfo{year}{2012}).

\bibitem[{\citenamefont{Choy et~al.}(2011)\citenamefont{Choy, Edge, Akhmerov,
  and Beenakker}}]{Choy11}
\bibinfo{author}{\bibfnamefont{T.-P.} \bibnamefont{Choy}},
  \bibinfo{author}{\bibfnamefont{J.~M.} \bibnamefont{Edge}},
  \bibinfo{author}{\bibfnamefont{A.~R.} \bibnamefont{Akhmerov}},
  \bibnamefont{and} \bibinfo{author}{\bibfnamefont{C.~W.~J.}
  \bibnamefont{Beenakker}}, \bibinfo{journal}{Phys. Rev. B}
  \textbf{\bibinfo{volume}{84}}, \bibinfo{pages}{195442}
  (\bibinfo{year}{2011}).

\bibitem[{\citenamefont{Nadj-Perge et~al.}(2013)\citenamefont{Nadj-Perge,
  Drozdov, Bernevig, and Yazdani}}]{NadjPerge13}
\bibinfo{author}{\bibfnamefont{S.}~\bibnamefont{Nadj-Perge}},
  \bibinfo{author}{\bibfnamefont{I.~K.} \bibnamefont{Drozdov}},
  \bibinfo{author}{\bibfnamefont{B.~A.} \bibnamefont{Bernevig}},
  \bibnamefont{and} \bibinfo{author}{\bibfnamefont{A.}~\bibnamefont{Yazdani}},
  \bibinfo{journal}{Phys. Rev. B} \textbf{\bibinfo{volume}{88}},
  \bibinfo{pages}{020407} (\bibinfo{year}{2013}).

\bibitem[{\citenamefont{Braunecker and Simon}(2013)}]{brauneckerPRL13}
\bibinfo{author}{\bibfnamefont{B.}~\bibnamefont{Braunecker}} \bibnamefont{and}
  \bibinfo{author}{\bibfnamefont{P.}~\bibnamefont{Simon}},
  \bibinfo{journal}{Phys. Rev. Lett.} \textbf{\bibinfo{volume}{111}},
  \bibinfo{pages}{147202} (\bibinfo{year}{2013}).

\bibitem[{\citenamefont{Pientka et~al.}(2013)\citenamefont{Pientka, Glazman,
  and von Oppen}}]{pientkaPRB13}
\bibinfo{author}{\bibfnamefont{F.}~\bibnamefont{Pientka}},
  \bibinfo{author}{\bibfnamefont{L.~I.} \bibnamefont{Glazman}},
  \bibnamefont{and} \bibinfo{author}{\bibfnamefont{F.}~\bibnamefont{von
  Oppen}}, \bibinfo{journal}{Phys. Rev. B} \textbf{\bibinfo{volume}{88}},
  \bibinfo{pages}{155420} (\bibinfo{year}{2013}).

\bibitem[{\citenamefont{Klinovaja et~al.}(2013)\citenamefont{Klinovaja, Stano,
  Yazdani, and Loss}}]{klinovajaPRL13}
\bibinfo{author}{\bibfnamefont{J.}~\bibnamefont{Klinovaja}},
  \bibinfo{author}{\bibfnamefont{P.}~\bibnamefont{Stano}},
  \bibinfo{author}{\bibfnamefont{A.}~\bibnamefont{Yazdani}}, \bibnamefont{and}
  \bibinfo{author}{\bibfnamefont{D.}~\bibnamefont{Loss}},
  \bibinfo{journal}{Phys. Rev. Lett.} \textbf{\bibinfo{volume}{111}},
  \bibinfo{pages}{186805} (\bibinfo{year}{2013}).

\bibitem[{\citenamefont{Vazifeh and Franz}(2013)}]{vazifehPRL13}
\bibinfo{author}{\bibfnamefont{M.~M.} \bibnamefont{Vazifeh}} \bibnamefont{and}
  \bibinfo{author}{\bibfnamefont{M.}~\bibnamefont{Franz}},
  \bibinfo{journal}{Phys. Rev. Lett.} \textbf{\bibinfo{volume}{111}},
  \bibinfo{pages}{206802} (\bibinfo{year}{2013}).

\bibitem[{\citenamefont{Pientka et~al.}(2014)\citenamefont{Pientka, Glazman,
  and von Oppen}}]{pientkaPRB14}
\bibinfo{author}{\bibfnamefont{F.}~\bibnamefont{Pientka}},
  \bibinfo{author}{\bibfnamefont{L.~I.} \bibnamefont{Glazman}},
  \bibnamefont{and} \bibinfo{author}{\bibfnamefont{F.}~\bibnamefont{von
  Oppen}}, \bibinfo{journal}{Phys. Rev. B} \textbf{\bibinfo{volume}{89}},
  \bibinfo{pages}{180505} (\bibinfo{year}{2014}).

\bibitem[{\citenamefont{R\"ontynen and Ojanen}(2014)}]{rontynenPRB14}
\bibinfo{author}{\bibfnamefont{J.}~\bibnamefont{R\"ontynen}} \bibnamefont{and}
  \bibinfo{author}{\bibfnamefont{T.}~\bibnamefont{Ojanen}},
  \bibinfo{journal}{Phys. Rev. B} \textbf{\bibinfo{volume}{90}},
  \bibinfo{pages}{180503} (\bibinfo{year}{2014}).

\bibitem[{\citenamefont{Kim et~al.}(2014)\citenamefont{Kim, Cheng, Bauer,
  Lutchyn, and Das~Sarma}}]{kimPRB14}
\bibinfo{author}{\bibfnamefont{Y.}~\bibnamefont{Kim}},
  \bibinfo{author}{\bibfnamefont{M.}~\bibnamefont{Cheng}},
  \bibinfo{author}{\bibfnamefont{B.}~\bibnamefont{Bauer}},
  \bibinfo{author}{\bibfnamefont{R.~M.} \bibnamefont{Lutchyn}},
  \bibnamefont{and}
  \bibinfo{author}{\bibfnamefont{S.}~\bibnamefont{Das~Sarma}},
  \bibinfo{journal}{Phys. Rev. B} \textbf{\bibinfo{volume}{90}},
  \bibinfo{pages}{060401} (\bibinfo{year}{2014}).

\bibitem[{\citenamefont{Brydon et~al.}(2015)\citenamefont{Brydon, Das~Sarma,
  Hui, and Sau}}]{brydonPRB15}
\bibinfo{author}{\bibfnamefont{P.~M.~R.} \bibnamefont{Brydon}},
  \bibinfo{author}{\bibfnamefont{S.}~\bibnamefont{Das~Sarma}},
  \bibinfo{author}{\bibfnamefont{H.-Y.} \bibnamefont{Hui}}, \bibnamefont{and}
  \bibinfo{author}{\bibfnamefont{J.~D.} \bibnamefont{Sau}},
  \bibinfo{journal}{Phys. Rev. B} \textbf{\bibinfo{volume}{91}},
  \bibinfo{pages}{064505} (\bibinfo{year}{2015}).

\bibitem[{\citenamefont{Yu}(1965)}]{yuActaPhysSin}
\bibinfo{author}{\bibfnamefont{L.}~\bibnamefont{Yu}}, \bibinfo{journal}{Acta
  Phys. Sin.} \textbf{\bibinfo{volume}{21}} (\bibinfo{year}{1965}).

\bibitem[{\citenamefont{Shiba}(1968)}]{shibaProgTheoPhys68}
\bibinfo{author}{\bibfnamefont{H.}~\bibnamefont{Shiba}},
  \bibinfo{journal}{Progress of Theoretical Physics}
  \textbf{\bibinfo{volume}{40}}, \bibinfo{pages}{435} (\bibinfo{year}{1968}).

\bibitem[{\citenamefont{Rusinov}(1968)}]{rusinovJETP69}
\bibinfo{author}{\bibfnamefont{A.~I.} \bibnamefont{Rusinov}},
  \bibinfo{journal}{Zh. Eksp. Teor. Fiz. Pisma. Red.}
  \textbf{\bibinfo{volume}{9}}, \bibinfo{pages}{146} (\bibinfo{year}{1968}).

\bibitem[{\citenamefont{Nadj-Perge et~al.}(2014)\citenamefont{Nadj-Perge,
  Drozdov, Li, Chen, Jeon, Seo, MacDonald, Bernevig, and
  Yazdani}}]{Nadj-Perge_Ferro_SC}
\bibinfo{author}{\bibfnamefont{S.}~\bibnamefont{Nadj-Perge}},
  \bibinfo{author}{\bibfnamefont{I.~K.} \bibnamefont{Drozdov}},
  \bibinfo{author}{\bibfnamefont{J.}~\bibnamefont{Li}},
  \bibinfo{author}{\bibfnamefont{H.}~\bibnamefont{Chen}},
  \bibinfo{author}{\bibfnamefont{S.}~\bibnamefont{Jeon}},
  \bibinfo{author}{\bibfnamefont{J.}~\bibnamefont{Seo}},
  \bibinfo{author}{\bibfnamefont{A.~H.} \bibnamefont{MacDonald}},
  \bibinfo{author}{\bibfnamefont{B.~A.} \bibnamefont{Bernevig}},
  \bibnamefont{and} \bibinfo{author}{\bibfnamefont{A.}~\bibnamefont{Yazdani}},
  \bibinfo{journal}{Science} \textbf{\bibinfo{volume}{346}},
  \bibinfo{pages}{602} (\bibinfo{year}{2014}).

\bibitem[{\citenamefont{{Sau} and {Brydon}}(2015)}]{brydonArxiv2014}
\bibinfo{author}{\bibfnamefont{J.~D.} \bibnamefont{{Sau}}} \bibnamefont{and}
  \bibinfo{author}{\bibfnamefont{P.~M.~R.} \bibnamefont{{Brydon}}},
  \bibinfo{journal}{ArXiv e-prints}  (\bibinfo{year}{2015}),
  \eprint{1501.03149}.

\bibitem[{\citenamefont{Nakosai et~al.}(2013)\citenamefont{Nakosai, Tanaka, and
  Nagaosa}}]{Nakosai13}
\bibinfo{author}{\bibfnamefont{S.}~\bibnamefont{Nakosai}},
  \bibinfo{author}{\bibfnamefont{Y.}~\bibnamefont{Tanaka}}, \bibnamefont{and}
  \bibinfo{author}{\bibfnamefont{N.}~\bibnamefont{Nagaosa}},
  \bibinfo{journal}{Phys. Rev. B} \textbf{\bibinfo{volume}{88}},
  \bibinfo{pages}{180503} (\bibinfo{year}{2013}).

\bibitem[{\citenamefont{Deutscher and Meunier}(1969)}]{Deutscher69}
\bibinfo{author}{\bibfnamefont{G.}~\bibnamefont{Deutscher}} \bibnamefont{and}
  \bibinfo{author}{\bibfnamefont{F.}~\bibnamefont{Meunier}},
  \bibinfo{journal}{Phys. Rev. Lett.} \textbf{\bibinfo{volume}{22}},
  \bibinfo{pages}{395} (\bibinfo{year}{1969}).

\bibitem[{\citenamefont{Hauser}(1969)}]{Hauser69}
\bibinfo{author}{\bibfnamefont{J.~J.} \bibnamefont{Hauser}},
  \bibinfo{journal}{Phys. Rev. Lett.} \textbf{\bibinfo{volume}{23}},
  \bibinfo{pages}{374} (\bibinfo{year}{1969}).

\bibitem[{\citenamefont{Gennes}(1966)}]{deGennes66}
\bibinfo{author}{\bibfnamefont{P.~D.} \bibnamefont{Gennes}},
  \bibinfo{journal}{Physics Letters} \textbf{\bibinfo{volume}{23}},
  \bibinfo{pages}{10 } (\bibinfo{year}{1966}), ISSN \bibinfo{issn}{0031-9163}.

\bibitem[{\citenamefont{Yamasaki et~al.}(2007)\citenamefont{Yamasaki, Sagayama,
  Goto, Matsuura, Hirota, Arima, and Tokura}}]{Yamasaki07}
\bibinfo{author}{\bibfnamefont{Y.}~\bibnamefont{Yamasaki}},
  \bibinfo{author}{\bibfnamefont{H.}~\bibnamefont{Sagayama}},
  \bibinfo{author}{\bibfnamefont{T.}~\bibnamefont{Goto}},
  \bibinfo{author}{\bibfnamefont{M.}~\bibnamefont{Matsuura}},
  \bibinfo{author}{\bibfnamefont{K.}~\bibnamefont{Hirota}},
  \bibinfo{author}{\bibfnamefont{T.}~\bibnamefont{Arima}}, \bibnamefont{and}
  \bibinfo{author}{\bibfnamefont{Y.}~\bibnamefont{Tokura}},
  \bibinfo{journal}{Phys. Rev. Lett.} \textbf{\bibinfo{volume}{98}},
  \bibinfo{pages}{147204} (\bibinfo{year}{2007}).

\bibitem[{\citenamefont{Yamasaki et~al.}(2008)\citenamefont{Yamasaki, Sagayama,
  Abe, Arima, Sasai, Matsuura, Hirota, Okuyama, Noda, and Tokura}}]{Yamasaki08}
\bibinfo{author}{\bibfnamefont{Y.}~\bibnamefont{Yamasaki}},
  \bibinfo{author}{\bibfnamefont{H.}~\bibnamefont{Sagayama}},
  \bibinfo{author}{\bibfnamefont{N.}~\bibnamefont{Abe}},
  \bibinfo{author}{\bibfnamefont{T.}~\bibnamefont{Arima}},
  \bibinfo{author}{\bibfnamefont{K.}~\bibnamefont{Sasai}},
  \bibinfo{author}{\bibfnamefont{M.}~\bibnamefont{Matsuura}},
  \bibinfo{author}{\bibfnamefont{K.}~\bibnamefont{Hirota}},
  \bibinfo{author}{\bibfnamefont{D.}~\bibnamefont{Okuyama}},
  \bibinfo{author}{\bibfnamefont{Y.}~\bibnamefont{Noda}}, \bibnamefont{and}
  \bibinfo{author}{\bibfnamefont{Y.}~\bibnamefont{Tokura}},
  \bibinfo{journal}{Phys. Rev. Lett.} \textbf{\bibinfo{volume}{101}},
  \bibinfo{pages}{097204} (\bibinfo{year}{2008}).

\bibitem[{\citenamefont{Kimura and Tokura}(2008)}]{Kimura08}
\bibinfo{author}{\bibfnamefont{T.}~\bibnamefont{Kimura}} \bibnamefont{and}
  \bibinfo{author}{\bibfnamefont{Y.}~\bibnamefont{Tokura}},
  \bibinfo{journal}{Journal of Physics: Condensed Matter}
  \textbf{\bibinfo{volume}{20}}, \bibinfo{pages}{434204}
  (\bibinfo{year}{2008}).

\bibitem[{\citenamefont{Murakawa et~al.}(2008)\citenamefont{Murakawa, Onose,
  Kagawa, Ishiwata, Kaneko, and Tokura}}]{Murakawa08}
\bibinfo{author}{\bibfnamefont{H.}~\bibnamefont{Murakawa}},
  \bibinfo{author}{\bibfnamefont{Y.}~\bibnamefont{Onose}},
  \bibinfo{author}{\bibfnamefont{F.}~\bibnamefont{Kagawa}},
  \bibinfo{author}{\bibfnamefont{S.}~\bibnamefont{Ishiwata}},
  \bibinfo{author}{\bibfnamefont{Y.}~\bibnamefont{Kaneko}}, \bibnamefont{and}
  \bibinfo{author}{\bibfnamefont{Y.}~\bibnamefont{Tokura}},
  \bibinfo{journal}{Phys. Rev. Lett.} \textbf{\bibinfo{volume}{101}},
  \bibinfo{pages}{197207} (\bibinfo{year}{2008}).

\bibitem[{\citenamefont{Sedlmayr et~al.}(2015)\citenamefont{Sedlmayr,
  Aguiar-Hualde, and Bena}}]{Sedlmayr15}
\bibinfo{author}{\bibfnamefont{N.}~\bibnamefont{Sedlmayr}},
  \bibinfo{author}{\bibfnamefont{J.~M.} \bibnamefont{Aguiar-Hualde}},
  \bibnamefont{and} \bibinfo{author}{\bibfnamefont{C.}~\bibnamefont{Bena}},
  \bibinfo{journal}{Phys. Rev. B} \textbf{\bibinfo{volume}{91}},
  \bibinfo{pages}{115415} (\bibinfo{year}{2015}).

\bibitem[{\citenamefont{Kajiwara et~al.}(2010)\citenamefont{Kajiwara, Harii,
  Takahashi, Ohe, Uchida, Mizuguchi, Umezawa, Kawai, Ando, Takanashi
  et~al.}}]{Kajiwara10}
\bibinfo{author}{\bibfnamefont{Y.}~\bibnamefont{Kajiwara}},
  \bibinfo{author}{\bibfnamefont{K.}~\bibnamefont{Harii}},
  \bibinfo{author}{\bibfnamefont{S.}~\bibnamefont{Takahashi}},
  \bibinfo{author}{\bibfnamefont{J.}~\bibnamefont{Ohe}},
  \bibinfo{author}{\bibfnamefont{K.}~\bibnamefont{Uchida}},
  \bibinfo{author}{\bibfnamefont{M.}~\bibnamefont{Mizuguchi}},
  \bibinfo{author}{\bibfnamefont{H.}~\bibnamefont{Umezawa}},
  \bibinfo{author}{\bibfnamefont{H.}~\bibnamefont{Kawai}},
  \bibinfo{author}{\bibfnamefont{K.}~\bibnamefont{Ando}},
  \bibinfo{author}{\bibfnamefont{K.}~\bibnamefont{Takanashi}},
  \bibnamefont{et~al.}, \bibinfo{journal}{Nature}
  \textbf{\bibinfo{volume}{464}}, \bibinfo{pages}{262} (\bibinfo{year}{2010}).

\bibitem[{\citenamefont{Johnson et~al.}(2013)\citenamefont{Johnson, Barone,
  Bombardi, Bean, Picozzi, Radaelli, Oh, Cheong, and Chapon}}]{Johnson13}
\bibinfo{author}{\bibfnamefont{R.~D.} \bibnamefont{Johnson}},
  \bibinfo{author}{\bibfnamefont{P.}~\bibnamefont{Barone}},
  \bibinfo{author}{\bibfnamefont{A.}~\bibnamefont{Bombardi}},
  \bibinfo{author}{\bibfnamefont{R.~J.} \bibnamefont{Bean}},
  \bibinfo{author}{\bibfnamefont{S.}~\bibnamefont{Picozzi}},
  \bibinfo{author}{\bibfnamefont{P.~G.} \bibnamefont{Radaelli}},
  \bibinfo{author}{\bibfnamefont{Y.~S.} \bibnamefont{Oh}},
  \bibinfo{author}{\bibfnamefont{S.-W.} \bibnamefont{Cheong}},
  \bibnamefont{and} \bibinfo{author}{\bibfnamefont{L.~C.}
  \bibnamefont{Chapon}}, \bibinfo{journal}{Phys. Rev. Lett.}
  \textbf{\bibinfo{volume}{110}}, \bibinfo{pages}{217206}
  (\bibinfo{year}{2013}).

\bibitem[{\citenamefont{Seki et~al.}(2012{\natexlab{a}})\citenamefont{Seki, Yu,
  Ishiwata, and Tokura}}]{Seki12}
\bibinfo{author}{\bibfnamefont{S.}~\bibnamefont{Seki}},
  \bibinfo{author}{\bibfnamefont{X.~Z.} \bibnamefont{Yu}},
  \bibinfo{author}{\bibfnamefont{S.}~\bibnamefont{Ishiwata}}, \bibnamefont{and}
  \bibinfo{author}{\bibfnamefont{Y.}~\bibnamefont{Tokura}},
  \bibinfo{journal}{Science} \textbf{\bibinfo{volume}{336}},
  \bibinfo{pages}{198} (\bibinfo{year}{2012}{\natexlab{a}}).

\bibitem[{\citenamefont{Seki et~al.}(2012{\natexlab{b}})\citenamefont{Seki,
  Ishiwata, and Tokura}}]{Seki12_2}
\bibinfo{author}{\bibfnamefont{S.}~\bibnamefont{Seki}},
  \bibinfo{author}{\bibfnamefont{S.}~\bibnamefont{Ishiwata}}, \bibnamefont{and}
  \bibinfo{author}{\bibfnamefont{Y.}~\bibnamefont{Tokura}},
  \bibinfo{journal}{Phys. Rev. B} \textbf{\bibinfo{volume}{86}},
  \bibinfo{pages}{060403} (\bibinfo{year}{2012}{\natexlab{b}}).

\bibitem[{\citenamefont{Schulz et~al.}(2012)\citenamefont{Schulz, Ritz, Bauer,
  Halder, Wagner, Franz, Pfleiderer, Everschor, Garst, and Rosch}}]{Schulz12}
\bibinfo{author}{\bibfnamefont{T.}~\bibnamefont{Schulz}},
  \bibinfo{author}{\bibfnamefont{R.}~\bibnamefont{Ritz}},
  \bibinfo{author}{\bibfnamefont{A.}~\bibnamefont{Bauer}},
  \bibinfo{author}{\bibfnamefont{M.}~\bibnamefont{Halder}},
  \bibinfo{author}{\bibfnamefont{M.}~\bibnamefont{Wagner}},
  \bibinfo{author}{\bibfnamefont{C.}~\bibnamefont{Franz}},
  \bibinfo{author}{\bibfnamefont{C.}~\bibnamefont{Pfleiderer}},
  \bibinfo{author}{\bibfnamefont{K.}~\bibnamefont{Everschor}},
  \bibinfo{author}{\bibfnamefont{M.}~\bibnamefont{Garst}}, \bibnamefont{and}
  \bibinfo{author}{\bibfnamefont{A.}~\bibnamefont{Rosch}},
  \bibinfo{journal}{Nat Phys} \textbf{\bibinfo{volume}{8}},
  \bibinfo{pages}{301} (\bibinfo{year}{2012}).

\bibitem[{\citenamefont{Nagaosa and Tokura}(2013)}]{Nagaosa13}
\bibinfo{author}{\bibfnamefont{N.}~\bibnamefont{Nagaosa}} \bibnamefont{and}
  \bibinfo{author}{\bibfnamefont{Y.}~\bibnamefont{Tokura}},
  \bibinfo{journal}{Nat Nano} \textbf{\bibinfo{volume}{8}},
  \bibinfo{pages}{899} (\bibinfo{year}{2013}).

\bibitem[{\citenamefont{Zhang and Zhang}(2009)}]{Zhang09}
\bibinfo{author}{\bibfnamefont{S.}~\bibnamefont{Zhang}} \bibnamefont{and}
  \bibinfo{author}{\bibfnamefont{S.~S.-L.} \bibnamefont{Zhang}},
  \bibinfo{journal}{Phys. Rev. Lett.} \textbf{\bibinfo{volume}{102}},
  \bibinfo{pages}{086601} (\bibinfo{year}{2009}).

\bibitem[{\citenamefont{Romming et~al.}(2013)\citenamefont{Romming, Hanneken,
  Menzel, Bickel, Wolter, von Bergmann, Kubetzka, and
  Wiesendanger}}]{Romming13}
\bibinfo{author}{\bibfnamefont{N.}~\bibnamefont{Romming}},
  \bibinfo{author}{\bibfnamefont{C.}~\bibnamefont{Hanneken}},
  \bibinfo{author}{\bibfnamefont{M.}~\bibnamefont{Menzel}},
  \bibinfo{author}{\bibfnamefont{J.~E.} \bibnamefont{Bickel}},
  \bibinfo{author}{\bibfnamefont{B.}~\bibnamefont{Wolter}},
  \bibinfo{author}{\bibfnamefont{K.}~\bibnamefont{von Bergmann}},
  \bibinfo{author}{\bibfnamefont{A.}~\bibnamefont{Kubetzka}}, \bibnamefont{and}
  \bibinfo{author}{\bibfnamefont{R.}~\bibnamefont{Wiesendanger}},
  \bibinfo{journal}{Science} \textbf{\bibinfo{volume}{341}},
  \bibinfo{pages}{636} (\bibinfo{year}{2013}).

\end{thebibliography}

\end{document}